\documentclass[aps,preprint,preprintnumbers,showpacs,showkeys]{revtex4}

\usepackage{graphicx}
\usepackage{bm}

\def\vec#1{\mbox{\boldmath$#1$}}

 \def\gsl#1{\rlap{\slash}#1}
\def\A{{\scriptsize\mbox{Zhu}}}
\def\B{{\scriptsize\mbox{Matheus {\it et al}.\ }}}
\def\C{{\scriptsize\mbox{Sugiyama {\it et al}.\ }}}
 
\begin{document}

\preprint{TH-953}

\title{Two-Hadron-Irreducible QCD Sum Rule for Pentaquark Baryon}

\author{Yoshihiko KONDO}
\email{kondo@kokugakuin.ac.jp}
\affiliation{%
Kokugakuin University, Higashi, Shibuya, Tokyo 150-8440, Japan
}
\author{Osamu MORIMATSU}
\email{osamu.morimatsu@kek.jp}
\author{Tetsuo NISHIKAWA}
\email{nishi@post.kek.jp} 
\affiliation{%
Institute of Particle and Nuclear Studies, 
High Energy Accelerator Research Organization, 1-1, Ooho, 
Tsukuba, Ibaraki, 305-0801, Japan
}

\date{\today}

\begin{abstract}
We point out that naive pentaquark correlations function include two-hadron-reducible contributions, which are given by convolution of baryon and meson correlation functions and have nothing to do with pentaquark.
We show that the two-hadron-reducible contributions are large in the operator product expansion of the correlation functions of three existing works on the pentaquark.
Therefore, it is dangerous to draw a conclusion from the sum rules using naive pentaquark correlation functions with naive ansatz for the spectral function under the dispersion integral.
Instead, we propose to use the two-hadron-irreducible correlation function, which is
obtained by subtracting the two-hadron-reducible contribution from the naive
correlation function.
Taking one of the works as an example we demonstrate how drastically the results can change if we remove the two-hadron-reducible part from the naive correlation function.
We obtain the result opposite to the original work for the parity of the pentaquark.
\end{abstract}
\maketitle

\newpage
Possible existence of an $S=+1$ exotic baryon state has recently been reported by LEPS collaboration in Spring-8 \cite{leps}.
In this experiment, a peak with the mass $\sim1540\,{\rm MeV}$ and the width
bounded by $25\,{\rm MeV}$ was observed in $K^{+}n$ channel from 
$\gamma n\rightarrow K^{+}K^{-}n$.
It was confirmed by subsequent experiments \cite{diana,clasa,clasb,saphir,itep,hermes,itep-2,zeus,clas-c}. 
In all cases the mass is near $1540\,{\rm MeV}$
and the width is small.

This state cannot be a three-quark state since it has $S=+1$, and the minimal quark content is $(uudd{\bar s})$.
It has come to be called \lq\lq pentaquark $\Theta^{+}$".
Other quantum numbers, spin, isospin and parity, have not yet been experimentally determined.
Besides  the calculation in the chiral quark soliton model \cite{diakonov}, which motivated the pentaquark search by LEPS collaboration, various theoretical approaches \cite{jaffe,capstick,karliner,sasaki,zhu,matheus,sugiyama} have been performed, but no consensus has yet been seen on these quantum numbers.

Clearly, the discoveries of exotic hadrons have opened a new field of strong interaction physics and will lead us to a deeper understanding of QCD.
In particular, what role confinement plays in exotic hadrons is an extremely interesting and important issue.

In this paper, we focus on the application of the QCD sum rule approach \cite{SVZ} to the pentaquark and discuss an issue which is characteristic for exotic hadrons. 
In QCD sum rule approach, the correlation function of interpolating fields is calculated by the use of 
the operator product expansion (OPE),
and is compared with the spectral representation via dispersion relation.
The sum rules relate hadron properties to the vacuum expectation values
of QCD operators, such as $\langle 0|{\bar q}q|0\rangle$ and $\langle 0|(\alpha_{s}/\pi)G^2|0\rangle$. 

Up to now,  three groups have reported results of QCD sum rules for the pentaquark \cite{zhu,matheus,sugiyama}.
Zhu \cite{zhu} estimated the mass of the pentaquark states with QCD sum rules and found that pentaquark states with isospin $I=0, 1, 2$ lie close to each other around $1.55 \pm 0.15 {\rm GeV}$.
He used the interpolating fields for the $I=0, 1, 2$ pentaquark states where three quarks and the remaining ${\bar s}q$ are both in a color adjoint representation.
Matheus {\it et al}.\ \cite{matheus} used a linear combination of two independent interpolating fields:  one is made of two scalar $ud$ diquarks and the other of two pseudo-scalar $ud$ diquarks, together with $\bar s$.
Their result for the pentaquark mass is $m_{\Theta^{+}}=1.55\pm 0.10\,{\rm GeV}$.
It should be noted, however, that their interpolating field has isospin $I=1$  although it was stated to have isospin $I=0$ in Ref.\cite{matheus}.
Sugiyama, Doi and Oka \cite{sugiyama} employed an interpolating field 
with $J=1/2$, $I=0$ and $S=+1$
constructed from color anti-triplet scalar and pseudoscalar $ud$ diquarks
and an $\bar s$ quark.
They derived sum rules for the positive and negative parity states.
They found that the pole residue obtained from the sum rule is positive for the negative parity state
but negative for the positive parity state,
which implies that the obtained negative parity state is a real one but the pole in the positive parity spectral function is spurious.
Thus, they concluded that the parity of the pentaquark is negative.
The mass $m_{\Theta^{+}}$ was estimated to be near $m_{\Theta^{+}}\sim 1.5\,{\rm GeV}$.

Although these three groups obtained the pentaquark mass close to the observed value,
we point out that their naive pentaquark correlation functions include two-hadron-reducible contributions, which are due to noninteracting propagation of the three-quark baryon and the meson and therefore have nothing to do with the pentaquark.
These contributions exist in the correlation function only for exotic hadrons and are potentially large.
Therefore, it is dangerous to draw a conclusion from the sum rules using naive pentaquark correlation functions with naive ansatz for the spectral function under the dispersion integral.
Instead, we propose to use the two-hadron-irreducible correlation function, which is
obtained by subtracting the two-hadron-reducible contribution from the naive
correlation function.
We show how large the two-hadron-reducible contributions are in the OPE of the correlation functions considered in the above three works.
Taking Ref.\cite{sugiyama} as an example, we then demonstrate how drastically the results change if we remove the two-hadron-reducible part from the naive correlation function.

In order to derive QCD sum rules for the pentaquark, one considers the correlation function
\begin{eqnarray}\label{cf}
  \Pi_P(p)=-i\int d^4x {\rm e}^{ipx}\langle T(\eta_P(x)\bar \eta_P(0))\rangle,
\end{eqnarray}
where $\eta_P(x)$ is the interpolating field for the pentaquark, a composite operator made of five quark fields.
The spectral function is given by the imaginary part of the correlation function and represents the physical spectrum generated by the interpolating field as
\begin{eqnarray}
  \rho_P(p)=\sum_n\delta(p-p_n)\langle 0|\eta_P(0)|n\rangle\langle n|\bar\eta_P(0)|0\rangle \qquad (p^0>0).
\end{eqnarray}
In the QCD sum rule approach the spectral function is usually parametrized by a pole plus continuum contribution, which is adopted also for the pentaquark in Refs.~\cite{zhu,matheus,sugiyama}:
\begin{eqnarray}
  \rho_P(p)=Z\delta(p^2-M^2_P)+\theta(p^2-s_{th})\rho_P^{OPE}(p^2),
\end{eqnarray}
where $M_P$ is the mass of the pentaquark, $s_{th}$ the effective continuum threshold energy squared and $\rho_P^{OPE}(p^2)$
the imaginary part of the correlation function in the OPE.
For ordinary hadrons this procedure is probably all right but for exotic hadrons it might be too naive as we discuss.

A remarkable feature of the pentaquark is that it can be decomposed into a color-singlet three-quark state, baryon, and a color-singlet quark-antiquark state, meson.
(Hereafter, we use the term baryon when its minimal quark-content is $qqq$.)
Therefore, the interpolating field for the pentaquark can be expressed as a sum of the product of  baryon and meson interpolating fields:
\begin{eqnarray}
  \eta_P(x)=\sum_i \eta^i_B(x)\eta^i_M(x)
\end{eqnarray}
where $\eta^i_B(x)$ and $\eta^i_M(x)$ are color-singlet baryon and meson interpolating fields, respectively:
$\eta_B \sim \epsilon_{abc}q^a q^b q^c$, $\eta_M\sim\bar q^e q^e$.

\begin{figure}
\begin{center}
\includegraphics{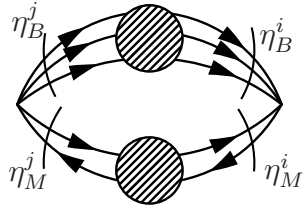}
\caption{Two-Hadron-Reducible (2HR) diagram.}
\label{2HR}
\end{center}
\end{figure}

\begin{figure}
\begin{center}
\includegraphics{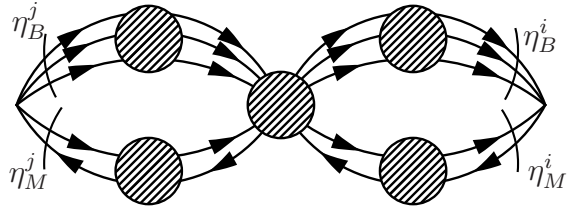}
\caption{Two-Hadron-Irreducible (2HI) diagram.}
\label{2HI}
\end{center}
\end{figure}

Due to this separability, the pentaquark correlation function, Eq.(\ref{cf}),  
has a part in which the baryon and the meson propagate independently without interacting each other.
We define this part as the two-hadron-reducible (2HR) part and the rest of the correlation function as the two-hadron-irreducible (2HI) part.
They are respectively given by
\begin{eqnarray}
  \langle T(\eta_P(x)\bar\eta_P(0))\rangle^{2HR}&=& 
  \sum_{ij}\langle T(\eta^i_B(x)\bar\eta^j_B(0))\rangle\langle T(\eta^i_M(x)\eta^{j*}_M(0))\rangle,\cr
  \langle T(\eta_P(x)\bar\eta_P(0))\rangle^{2HI} &=& \langle T(\eta_P(x)\bar\eta_P(0))\rangle
  -\sum_{ij}\langle T(\eta^i_B(x)\bar\eta^j_B(0))\rangle\langle T(\eta^i_M(x)\bar\eta^{j*}_M(0))\rangle
\end{eqnarray}
and
\begin{eqnarray}
  \Pi_P^{2HR}(p)&=&i\sum_{ij}\int {d^4q \over (2\pi)^4}\Pi^{ij}_B(q)\Pi^{ij}_M(p-q),\cr
  \Pi_P^{2HI}(p)&=& \Pi_P(p)-i\sum_{ij}\int {d^4q \over (2\pi)^4}\Pi^{ij}_B(q)\Pi^{ij}_M(p-q),
\end{eqnarray}
where $\Pi^{ij}_B(p)$ and $\Pi^{ij}_M(p)$ are baryon and meson correlation functions:
\begin{eqnarray}
  \Pi^{ij}_B(p)&=&-i\int d^4x {\rm e}^{ipx}\langle T(\eta^i_B(x)\bar \eta^j_B(0))\rangle,\\
  \Pi^{ij}_M(p)&=&-i\int d^4x {\rm e}^{ipx}\langle T(\eta^i_M(x)\eta^{j*}_M(0))\rangle.
\end{eqnarray}
Diagrammatically, the 2HR and 2HI parts of the correlation function are represented as Figures \ref{2HR} and \ref{2HI}, respectively.
Clearly, the 2HR part of the pentaquark correlation function is completely determined by the baryon and meson correlation functions and has nothing to with the pentaquark.

Some comments are in order here.
The 2HR and 2HI parts have the same analytic property as the total correlation function.
The concept of the two-hadron-reducible (2HR) and two-hadron-irreducible (2HI) contributions in the correlation function of the composite particle is a generalization of the two-particle-reducible (2PR) and two-particle-irreducible (2PI) contributions in the correlation function of the elementary particle.

Let us next look at the separation of the 2HR and 2HI parts in the spectral function.
We suppose that the lowest states generated by $\eta_B$ and $\eta_M$ are spin-1/2 baryon $B$ and spin-0 meson $M$, respectively.
Consider only the contribution of the $BM$ scattering states in the spectral function just for simplicity, which is given by
\begin{eqnarray}
 \rho^{BM}_P(p)
&=&\sum_s\int d^3k d^3q (2\pi)^3\delta^4(p-k-q) \langle0|\eta_P(0)|k qs out\rangle \langle k qs  out|\bar\eta_P(0)|0\rangle\quad (p^0>0),
\label{rhoBM-1}
\end{eqnarray}
where $q$ and $k$ are the momentum of the baryon and the meson, respectively, and $s$ the spin projection of the baryon. $|k qs out\rangle$ denote the baryon-meson scattering states with {\it out} boundary condition.
One can relate the matrix elements in Eq.(\ref{rhoBM-1}) with the $T$-matrix for the $BM$ scattering by means of the reduction formula as
\begin{eqnarray}
&&\langle k qs out|\bar\eta(0)|0\rangle
=\lambda_B\lambda_M{1\over (2\pi)^32\sqrt{\omega_k E_q}}\cr
&&\times\bar u(qs)
\left\{1-i\int{d^4q'\over(2\pi)^4}T(k,q,k+q-q',q')
{1\over (k+q-q')^2-m^2+i\eta}{1 \over \gsl q'-M+i\eta}\right\},
\label{me-1}
\\
&&\langle 0|\eta(0)|k qs out\rangle
=\lambda^*_B\lambda^*_M{1\over (2\pi)^32\sqrt{\omega_k E_q}}\cr
&&\times\left\{1+i\int{d^4q'\over(2\pi)^4}{1\over (k+q-q')^2-m^2-i\eta}{1 \over \gsl q'-M-i\eta}T^{\dagger}(k,q,k+q-q',q')\right\}u(qs),
\label{me-2}
\end{eqnarray}
where $\omega_k=\sqrt{{\bf k}^2 +m^2}$, $E_q=\sqrt{{\bf q}^2 +M^2}$, $\lambda_{B(M)}$ is the coupling strength of the interpolating field to the baryon (meson) state and $M (m)$ is the baryon (meson) mass.
$T(k,q,k',q')$ is the $T$-matrix for the process $kq\rightarrow k'q'$.
By substituting Eqs.(\ref{me-1}) and (\ref{me-2}) into Eq.(\ref{rhoBM-1}), we obtain 
\begin{eqnarray}
\rho^{BM}_P(p)&=&-{1\over \pi}|\lambda_B|^2|\lambda_M|^2{\rm Im}\left\{i\int {d^4q \over (2\pi)^4}{1 \over (p-q)^2-m^2+i\eta}{1 \over \gsl q - M+i\eta}\right.\cr
&&-\int {d^4q\over (2\pi)^4}{d^4q'\over (2\pi)^4}{1 \over (p-q)^2-m^2+i\eta} {1 \over \gsl q - M+i\eta}\cr
&&\left.\times T(p-q,q,p-q',q'){1 \over \gsl q' - M+i\eta}{1 \over (p-q')^2-m^2+i\eta}\right\}.
\label{rhoBM-2}
\end{eqnarray}
The first term of Eq.(\ref{rhoBM-2}) is the 2HR contribution due to the trivial noninteracting contribution of the $BM$ intermediate states and the second term is the 2HI contribution.
If the pentaquark is a bound state in the $BM$ channel the spectral function has an additional contribution from the bound state, while if the pentaquark is a resonance in the $BM$ channel the effect of the pentaquark lies in the $BM$ $T$-matrix as a pole at a complex pentaquark energy.
In any case, the 2HR contribution is not related to the pentaquark.

Some further comments are in order here.
2HR contributions discussed here exist commonly in the correlation functions for exotic hadrons but not for ordinary hadrons.
Crucial assumption here is confinement.
Namely, we assume that only color-singlet states contribute to the spectral function.
Therefore, the separability of the pentaquark into color-singlet baryon and meson is the origin of the existence of the 2HR contribution.
Another point to be mentioned is that by removing the 2HR part we do not take out all the baryon-meson continuum contributions.
We just take out noninteracting contributions.
The continuum contributions included in the 2HI part cannot be separated from the pentaquark contribution.
The 2HI part of the spectral function does not have positivity because of subtraction, but the positivity is not necessary to construct sum rules.

Let us turn to the separation of the 2HR and 2HI parts in the OPE.
There are various ways of constructing the interpolating field for the pentaquark.
So far, three ways of the construction have been proposed.
Zhu~\cite{zhu} used the following interpolating field for the $I=0$ pentaquark:
\begin{eqnarray}
\eta_\A
&=&{1 \over \sqrt{2}}\epsilon_{abc}(u^{Ta}C\gamma_5d^b)
     [u^g(\bar s^gi\gamma_5d^c)-(u \leftrightarrow d)],
     \label{current-zhu}
\end{eqnarray}
where $a$, $b$, $c,\ldots$ are color indices and $C=i\gamma_2 \gamma_0$.
Mathues {\it et al}.~\cite{matheus}\ and Sugiyama {\it et al}.~\cite{sugiyama}\
considered the interpolating fields based on the conjecture by Jaffe and Wilczek~\cite{jaffe}:
\begin{eqnarray}
\eta_\B&=&t\eta_1+\eta_2,\cr
\eta_1
&=&{1 \over \sqrt{2}}\epsilon_{abc}(u^{Ta}C\gamma_5d^b)(u^{Tc}C\gamma_5d^e)C(\bar s^e)^T-(u \leftrightarrow d),\cr
\eta_2
  &=&{1 \over \sqrt{2}}\epsilon_{abc}(u^{Ta}Cd^b)(u^{Tc}C\gamma_5d^c)C(\bar s^g)^T-(u \leftrightarrow d),\\
  \label{current-matheus}
\eta_\C  &=&\epsilon_{abc}\epsilon_{def}\epsilon_{cfg}(u^{Ta}Cd^b)(u^{Td}C\gamma_5d^e)C(\bar s^g)^T.
\label{current-sugiyama}
\end{eqnarray}
We first note that after Fierz transformation the interpolating field can be expressed as a sum of the product of baryon and meson interpolating fields.
The interpolating field used by Zhu can be written as
\begin{eqnarray}
\eta_\A
&=&{1 \over4\sqrt{2}}\epsilon_{abc}(u^{Ta}C\gamma_5d^b)\left\{-d^c(\bar s i\gamma_5 u)
  -\gamma^\mu d^c(\bar s i\gamma_5\gamma_\mu u)\right.\cr
  &&\left.-\frac{1}{2}\sigma^{\mu\nu}d^c(\bar s i\sigma_{\mu\nu}\gamma_5 u)
  -\gamma^\mu\gamma_5 d^c(\bar s i\gamma_\mu u)
  -\gamma_5 d^c(\bar s iu)-(u \leftrightarrow d)\right\}.
\end{eqnarray}
The interpolating fields considered by Matheus {\it et al}.\  and Sugiyama {\it et al}.\ 
are also written as follows, respectively:
\begin{eqnarray}
  \eta_\B&=&t\eta_1+\eta_2\cr
  \eta_1
  &=&{1 \over \sqrt{2}}\epsilon_{abc}(u^{Ta}C\gamma_5d^b)(u^{Tc}C\gamma_5d^e)C(\bar s^e)^T-(u \leftrightarrow d)\cr
  &=&{1 \over \sqrt{2}}{1\over 4}\left\{
  -[\epsilon_{abc}(u^{Ta}C\gamma_5d^b)u^c](\bar s\gamma_5 d)
  -[\epsilon_{abc}(u^{Ta}C\gamma_5d^b)\gamma^\mu u^c](\bar s\gamma_5\gamma_\mu d)\right.\cr
  &&+{1\over2}[\epsilon_{abc}(u^{Ta}C\gamma_5d^b)\sigma^{\mu\nu}u^c](\bar s\sigma_{\mu\nu}\gamma_5 d)
  +[\epsilon_{abc}(u^{Ta}C\gamma_5d^b)\gamma^\mu\gamma_5 u^c](\bar s\gamma_\mu d)\cr
  &&\left.-[\epsilon_{abc}(u^{Ta}C\gamma_5d^b)\gamma_5 u^c](\bar sd)\right\}-(u \leftrightarrow d)
\cr
  \eta_2
  &=&{1 \over \sqrt{2}}\epsilon_{abc}(u^{Ta}Cd^b)(u^{Tc}Cd^c)C(\bar s^g)^T-(u \leftrightarrow d)\cr
  &=&{1 \over \sqrt{2}}{1\over 4}\left\{-[\epsilon_{abc}(u^{Ta}Cd^b)u^c](\bar sd)
  +[\epsilon_{abc}(u^{Ta}Cd^b)\gamma^\mu u^c](\bar s\gamma_\mu d)\right.\cr
  &&+{1\over2}[\epsilon_{abc}(u^{Ta}Cd^b)\sigma^{\mu\nu}u^c](\bar s\sigma_{\mu\nu}d)
  +[\epsilon_{abc}(u^{Ta}Cd^b)\gamma^\mu\gamma_5 u^c](\bar s\gamma_5\gamma_\mu d)\cr
  &&\left.-[\epsilon_{abc}(u^{Ta}Cd^b)\gamma_5 u^c](\bar s\gamma_5d)\right\}-(u \leftrightarrow d),\\
  \eta_\C 
    &=&{1\over 4}\epsilon_{abc}(u^{Ta}Cd^b)\left\{-d^c(\bar s\gamma_5 u)
  -\gamma^\mu d^c(\bar s\gamma_5\gamma_\mu u)\right.\cr
  &&\left.+\frac{1}{2}\sigma^{\mu\nu}d^c(\bar s\sigma_{\mu\nu}\gamma_5 u)
  +\gamma^\mu\gamma_5 d^c(\bar s\gamma_\mu u)
  -\gamma_5 d^c(\bar su)-(u \leftrightarrow d)\right\}.
\end{eqnarray}
The 2HR parts in the correlation functions are obtained
by contracting a quark field in $\eta^i_B(x)$ with that in $\bar\eta^j_B(0)$
and contracting a quark field in $\eta^i_M(x)$ with that in $\eta^{j*}_M(0)$,
while the terms including contraction of a quark field in the baryon with that in the meson belong to the 2HI part.
We show the total and the 2HI part of the spectral function in the OPE.
For the correlation functions considered by Zhu, the results are
\begin{eqnarray}\label{Zhuspactral}
  \rho_\A(p)&=&
  \left\{{(p^2)^5 \over 2^{20} 3\cdot5\cdot7\pi^8}
  -{11(p^2)^3 \over 2^{14} 3^2 5 \pi^6}m_s\langle \bar q q\rangle
  +{(p^2)^3 \over 2^{16} 3 \pi^6}m_s\langle \bar s s\rangle\right.
  \cr&&\left.+{23(p^2)^3 \over 2^{19} 3^2 5 \pi^6}\langle {\alpha_s\over\pi}G^2\rangle
  +{5(p^2)^3 \over 2^{11} 3^2 \pi^4}\langle \bar qq \bar qq\rangle
  +{11(p^2)^3 \over 2^{12} 3^2 \pi^4}\langle \bar qq\bar ss\rangle
\right\}\gsl p\cr
  &&+{(p^2)^5 \over 2^{20} 3^2 5^2 \pi^8}m_s
  -{7 (p^2)^4 \over 2^{16} 3^2 5 \pi^6}\langle \bar q q\rangle
  -{(p^2)^4 \over 2^{17} 3^2 5 \pi^6}\langle \bar s s\rangle
  \cr&&+{25(p^2)^3 \over 2^{18} 3^2 \pi^6}\langle \bar q g_s\sigma\cdot G q\rangle
  -{7(p^2)^3 \over 2^{18} 3^2 \pi^6}\langle \bar s g_s\sigma\cdot G s\rangle,
\cr 
 \rho^{2HI}_\A(p)&=&
  \left\{
  -{(p^2)^3 \over 2^{15} 3^2 5 \pi^6}m_s\langle \bar q q\rangle
  +{(p^2)^3 \over 2^{16} 3^2 5 \pi^6}\langle {\alpha_s\over\pi}G^2\rangle
  -{(p^2)^3 \over 2^{13} 3^2 \pi^4}\langle \bar qq \bar qq\rangle
  +{(p^2)^3 \over 2^{13} 3^2 \pi^4}\langle \bar qq\bar ss\rangle
\right\}\gsl p\cr
  &&+{(p^2)^5 \over 2^{20} 3^2 5^2 \pi^8}m_s
  +{(p^2)^4 \over 2^{17} 3^2 5 \pi^6}\langle \bar q q\rangle
  -{(p^2)^4 \over 2^{17} 3^2 5 \pi^6}\langle \bar s s\rangle
  \cr&&-{5(p^2)^3 \over 2^{18} 3^2 \pi^6}\langle \bar q g_s\sigma\cdot G q\rangle
  -{7(p^2)^3 \over 2^{18} 3^2 \pi^6}\langle \bar s g_s\sigma\cdot G s\rangle,
\end{eqnarray}
where we have included the condensate $\langle {\alpha_s\over\pi}G^2\rangle$
, which is missing in Ref.\cite{zhu}.
For the correlation functions considered by Matheus {\it et al}.\  and Sugiyama {\it et al}.\  the results are
\begin{eqnarray}
\rho_\B(p)&=&
\left\{c_1\frac{(p^2)^5}{2^{18} 3^2 5^2 7\pi^8} 
+c_1\frac{(p^2)^3}{2^{14} 3^2 5\pi^6}m_s \langle{\bar s}s\rangle 
+c_2\frac{(p^2)^3}{2^{17} 3^2 5\pi^6}\langle {\alpha_s\over\pi}G^2\rangle
\right.
\cr
&&
\left.
-c_1\frac{(p^2)^2}{2^{15} 3^2 \pi^6}m_s \langle{\bar s}g\sigma\cdot Gs\rangle 
+c_3\frac{(p^2)^2 }{2^{10} 3^{2} \pi^4}\langle{\bar q}q{\bar q}q\rangle
\right\}\gsl p\cr
&&
+c_1\frac{(p^2)^5}{2^{18} 3^2 5^2 \pi^8}m_{s}
-c_1\frac{(p^2)^4}{2^{15} 3^2 5\pi^6}\langle{\bar s}s\rangle 
+c_1\frac{(p^2)^3}{2^{15} 3^2\pi^6}\langle{\bar s}g\sigma\cdot Gs\rangle ,\cr
\rho^{2HI}_\B(p)&=&
\left\{c_1\frac{(p^2)^5}{2^{20} 3^2 5^2 7\pi^8} 
+c_1\frac{(p^2)^3}{2^{16} 3^2 5\pi^6}m_s \langle{\bar s}s\rangle 
-c_4\frac{(p^2)^3}{2^{19} 3^2 5\pi^6}\langle {\alpha_s\over\pi}G^2\rangle
\right.
\cr
&&
\left.
-c_1\frac{(p^2)^2}{2^{17} 3^2 \pi^6}m_s \langle{\bar s}g\sigma\cdot Gs\rangle 
+c_3\frac{(p^2)^2 }{2^{12} 3^{2} \pi^4}\langle{\bar q}q{\bar q}q\rangle
\right\}\gsl p\cr
&&
+c_1\frac{(p^2)^5}{2^{20} 3^2 5^2 \pi^8}m_{s}
-c_1\frac{(p^2)^4}{2^{17} 3^2 5\pi^6}\langle{\bar s}s\rangle 
+c_1\frac{(p^2)^3}{2^{17} 3^2\pi^6}\langle{\bar s}g\sigma\cdot Gs\rangle,
\end{eqnarray}
where $c_1 =5t^2 +2t+5$, $c_2 =(t-1)^2$, $c_3 =7t^2 -2t-5$ and $c_{4}=11t^2 +14t+11$,
\begin{eqnarray}\label{SDKspectral}
  \rho_\C(p)&=&
  \left\{{(p^2)^5 \over 2^{16} 3^2 5^2 7\pi^8}
  +{(p^2)^3 \over 2^{12} 3^2 5 \pi^6}m_s\langle \bar s s\rangle
  +{(p^2)^3 \over 2^{14} 3^2 5 \pi^6}\langle {\alpha_s\over\pi}G^2\rangle
  \right.\cr
  &&\left.
  -{p^4 \over 2^{13} 3^2 \pi^6}m_s\langle \bar s g_s\sigma\cdot G s\rangle\right\}\gsl p\cr
  &&+{(p^2)^5 \over 2^{16} 3^2 5^2 \pi^8}m_s
  -{(p^2)^4 \over 2^{13} 3^2 5 \pi^6}\langle \bar s s\rangle
  +{(p^2)^3 \over 2^{13} 3^2 \pi^6}\langle \bar s g_s\sigma\cdot G s\rangle ,\cr
 \rho^{2HI}_\C(p)&=&
  \left\{-{(p^2)^5 \over 2^{19} 3^2 5^2\pi^8}
  -{7 (p^2)^3 \over 2^{15} 3^2 5 \pi^6}m_s\langle \bar s s\rangle
  +{(p^2)^3 \over 2^{18} 3^2 5 \pi^6}\langle {\alpha_s\over\pi}G^2\rangle\right.\cr
  &&\left.
  +{7 p^4 \over 2^{16} 3^2 \pi^6}m_s\langle \bar s g_s\sigma\cdot G s\rangle\right\}\gsl p\cr
  &&-{7 (p^2)^5 \over 2^{19} 3^2 5^2 \pi^8}m_s
  +{7 (p^2)^4 \over 2^{16} 3^2 5 \pi^6}\langle \bar s s\rangle
  -{7 (p^2)^3 \over 2^{16} 3^2 \pi^6}\langle \bar s g_s\sigma\cdot G s\rangle.
\end{eqnarray}
As can be seen, the 2HR part of the correlation function is large in general at least of the same order as the 2HI part.
In particular, for the interpolating field, Eq.(\ref{current-sugiyama}), the Wilson coeffcients of the operators in the 2HR part are $-15/7$ of the 2HI part up to dimension 6 except for the operator, ${\alpha_s\over\pi}G^2$.

\begin{figure}[h]
\begin{center}
\rotatebox{90}
{\includegraphics[width=4.0 in]{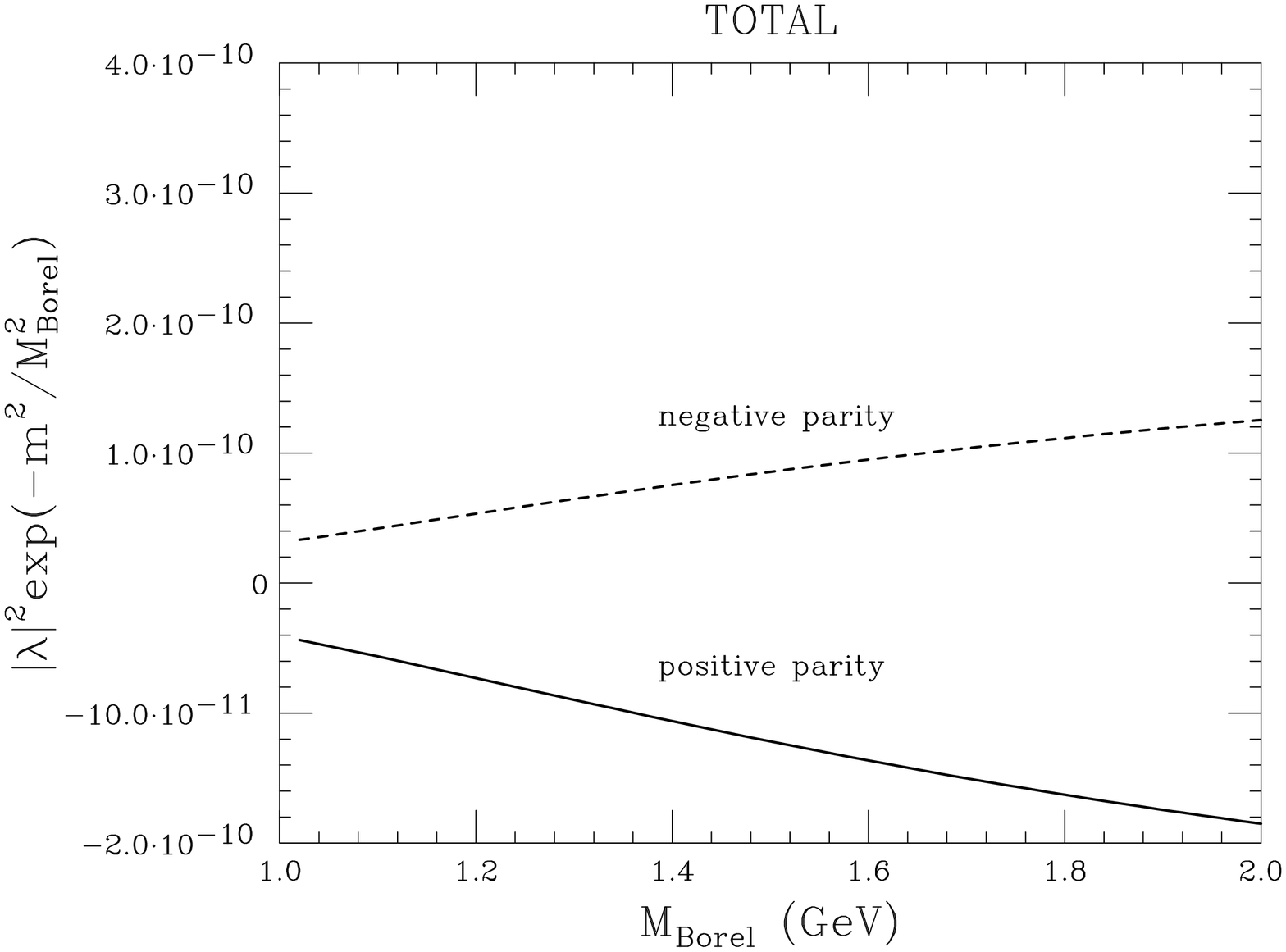}}
\caption{$|\lambda|^2\exp(-M_P^2/M_{\rm Borel}^2)$ vs. $M_{\rm Borel}$ for the total spectral function with $\sqrt{s_{th}}=1.8\,{\rm GeV}$}
\label{TOTAL-SR}
\vskip 0.2 in
\rotatebox{90}
{\includegraphics[width=4.0 in]{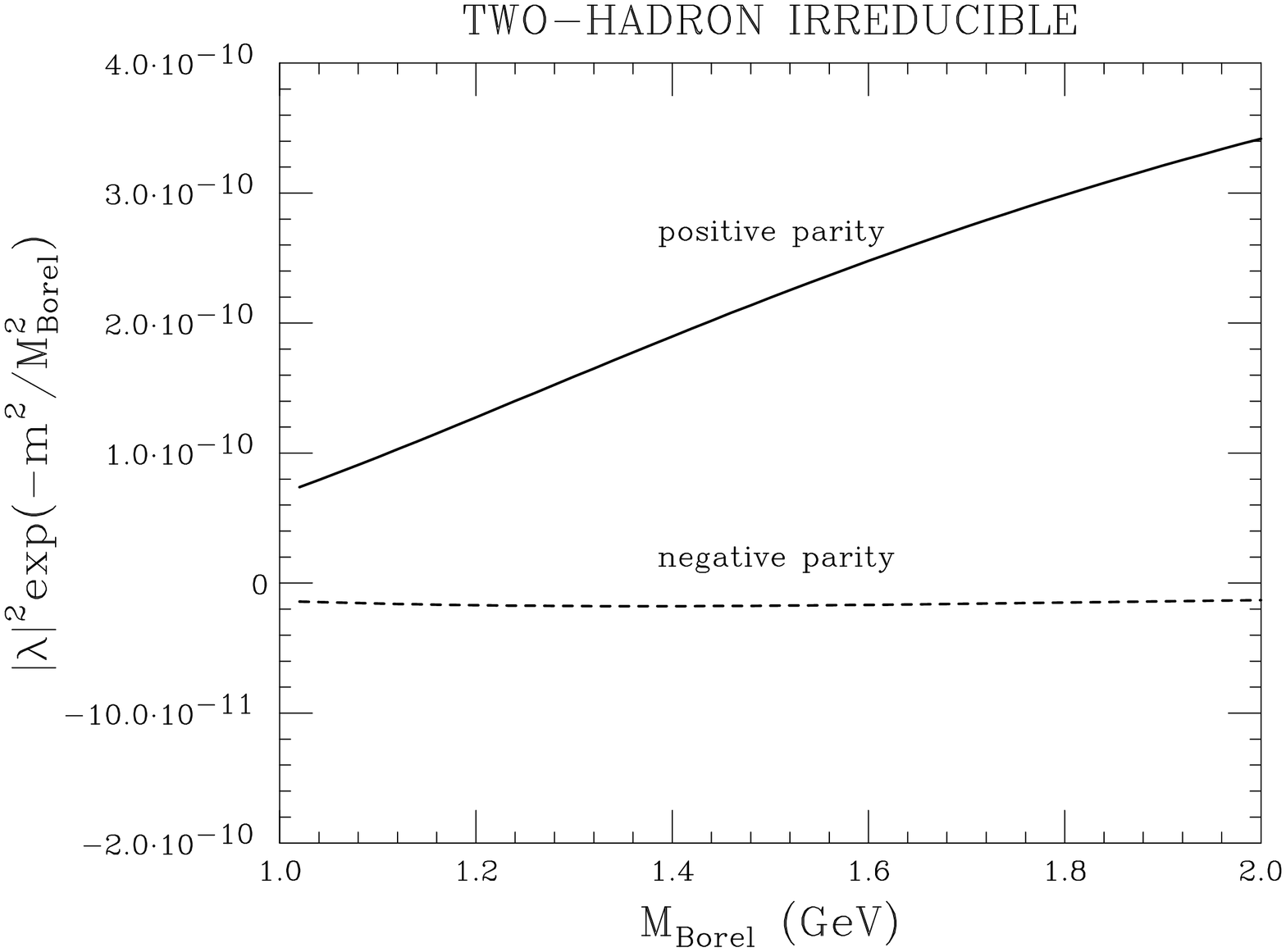}}
\caption{$|\lambda|^2\exp(-M_P^2/M_{\rm Borel}^2)$ vs. $M_{\rm Borel}$ for the 2HI spectral function with $\sqrt{s_{th}}=1.8\,{\rm GeV}$}
\label{2HI-SR}
\end{center}
\end{figure}

Now,  taking the interpolating field of Sugiyama {\it et al}.\ \cite{sugiyama}, Eq.(\ref{current-sugiyama}), we will demonstrate how drastically the results of the sum rule can change if we remove the 2HR part from the total correlation function.
First, we explain the results of Ref.\cite{sugiyama}.
They defined the positive-parity and negative-parity spectral functions by  
\begin{eqnarray}\label{SF1}
\rho_{\pm}(p_0)={1\over4}Tr[(\gamma_0\pm 1)\rho(p_0, \vec p=0)].
\end{eqnarray}
By parametrizing the spectral function by a pole plus continuum contribution as
\begin{eqnarray}\label{SF2}
  \rho_{\pm}(p_0)=|\lambda_\pm|^2\delta(p_0-m_{\pm})+\theta(p_0-\sqrt{s_{th}})\rho_P^{OPE}(p_0),
\end{eqnarray}
they derived sum rules,
\begin{eqnarray}\label{SR1}
&&|\lambda_\pm|^2\exp\left(-{m_\pm^2\over M_{\rm Borel}^2}\right)
=\int_0^{\sqrt{s_{th}}}dp_0\rho^{OPE}_{\pm}(p_0)\exp\left(-{p_0^2\over M_{\rm Borel}^2}\right),\\
\label{SR2}
&&|\lambda_\pm|^2m_{\pm}^2\exp\left(-{m_\pm^2\over M_{\rm Borel}^2}\right)=\int_0^{\sqrt{s_{th}}}dp_0\rho^{OPE}_{\pm}(p_0)p_0^2\exp\left(-{p_0^2\over M_{\rm Borel}^2}\right),
\end{eqnarray}
where $\sqrt{s_{th}}$ is the effective continuum threshold energy and the parameter of the weight function. $M_{\rm Borel}$ is called the Borel mass.
Fig.\ref{TOTAL-SR} shows the right-hand side of Eq.(\ref{SR1}) as a function of $M_{\rm Borel}$ with the standard values of the quark mass and QCD condensates, $m_s=0.12\,{\rm GeV}$, $\langle\bar ss\rangle=-0.8\times(0.23\,{\rm GeV})^3$, $\langle{\alpha_s\over\pi}G^2\rangle=(0.33\,{\rm GeV})^4$, $m_0^2=0.8\,{\rm GeV}^2$, where $m_0^2\equiv \langle\bar qg_s\sigma\cdot Gq\rangle/\langle\bar qq\rangle$ and $\sqrt{s_{th}}$ is taken to be $1.8\,{\rm GeV}$.
The solid and dashed lines are for the positive and negative-parity states, respectively.
As can be seen, the right-hand side of Eq.(\ref{SR1}) is positive for the negative-parity state but negative for the positive-parity state.
Since the left-hand side of Eq.(\ref{SR1}) must be positive, it was concluded that the obtained negative parity state is a real one but the pole in the positive parity spectral function is spurious.
The mass of the $\Theta^+$ was also predicted to be $\sim$ 1.5 GeV from the ratio of Eq.(\ref{SR2}) and Eq.(\ref{SR1}).

Though we have not been fully convinced if the procedure of Ref.\cite{sugiyama} is legitimate, we temporarily accept it in order to focus on the effect of removing the 2HR contributions on the results of sum rules.
We replace the total spectral function by the 2HI part but assume the same form, Eq.(\ref{SF2}), for the spectral function under the integral, where $\rho_P^{OPE}$ in the continuum contribution is also replaced by the 2HI part.
Fig.\ref{2HI-SR} is the same figure as Fig.\ref{TOTAL-SR} after replacing the total spectral function by the 2HI part.
Now, the right-hand side of Eq.(\ref{SR1}) is positive for the positive-parity state but negative for the negative-parity state.
This result was expected because the Wilson coeffcients of the operators in the 2HR part are $-15/7$ of those in the 2HI part up to dimension 6 except for the operator, ${\alpha_s\over\pi}G^2$.
Therefore, the sum rule for the 2HI part of the spectral function leads us to the opposite conclusion that  the obtained positive parity state is a real one but the pole in the negative parity spectral function is spurious.
The mass of the $\Theta^+$ is obtained to be $\sim 1.4\,{\rm GeV}$.
This result shows that it is crucial to remove the 2HR part from the total correlation function in order to correctly extract information on the pentaquark.
We, however, would like to suspend the conclusion that the parity of the pentaquark is positive until we make sure that all the procedures of deriving sum rules are reasonable.
We will report the results of our reinvestigation in near future \cite{kondo}.

Some final comments are in order here.
Logically, there is nothing wrong to use the total correlation function.
In order to derive sum rules, however, if the correlation function has a large background contribution which has nothing to do with what one wants to extract information about, pentaquark in the present case, it would be extremely difficult unless one knows very precise information about the background.
It is much better if the background can be exactly separated, which is what we proposed in the present paper.
Also, we would like to make a brief comment on the Lattice study of the pentaquark.
In the lattice calculation, the mass of the pentaquark is extracted from the long-time behavior of the total correlation function, which contains large two-hadron-reducible contribution.
It is necessary to remove this two-hadron-reducible contribution in order to obtain information on the pentaquark baryon\cite{sasaki}.
It would be interesting to see the relevance of the present work in the context of the Lattice study of the pentaquark.

We would like to thank Y.~Akaishi, T.~Doi, Y.~Kanada-E'nyo, M.~Oka and K.~Yazaki for useful discussion.
\def\Ref#1{[\ref{#1}]}
\def\Refs#1#2{[\ref{#1},\ref{#2}]}
\def\npb#1#2#3{{Nucl. Phys.\,}{\bf B{#1}}\,(#3)\,#2}
\def\npa#1#2#3{{Nucl. Phys.\,}{\bf A{#1}}\,(#3)\,#2}
\def\np#1#2#3{{Nucl. Phys.\,}{\bf{#1}}\,(#3)\,#2}
\def\plb#1#2#3{{Phys. Lett.\,}{\bf B{#1}}\,(#3)\,#2}
\def\prl#1#2#3{{Phys. Rev. Lett.\,}{\bf{#1}}\,(#3)\,#2}
\def\prd#1#2#3{{Phys. Rev.\,}{\bf D{#1}}\,(#3)\,#2}
\def\prc#1#2#3{{Phys. Rev.\,}{\bf C{#1}}\,(#3)\,#2}
\def\prb#1#2#3{{Phys. Rev.\,}{\bf B{#1}}\,(#3)\,#2}
\def\pr#1#2#3{{Phys. Rev.\,}{\bf{#1}}\,(#3)\,#2}
\def\ap#1#2#3{{Ann. Phys.\,}{\bf{#1}}\,(#3)\,#2}
\def\prep#1#2#3{{Phys. Reports\,}{\bf{#1}}\,(#3)\,#2}
\def\rmp#1#2#3{{Rev. Mod. Phys.\,}{\bf{#1}}\,(#3)\,#2}
\def\cmp#1#2#3{{Comm. Math. Phys.\,}{\bf{#1}}\,(#3)\,#2}
\def\ptp#1#2#3{{Prog. Theor. Phys.\,}{\bf{#1}}\,(#3)\,#2}
\def\ib#1#2#3{{\it ibid.\,}{\bf{#1}}\,(#3)\,#2}
\def\zsc#1#2#3{{Z. Phys. \,}{\bf C{#1}}\,(#3)\,#2}
\def\zsa#1#2#3{{Z. Phys. \,}{\bf A{#1}}\,(#3)\,#2}
\def\intj#1#2#3{{Int. J. Mod. Phys.\,}{\bf A{#1}}\,(#3)\,#2}
\def\sjnp#1#2#3{{Sov. J. Nucl. Phys.\,}{\bf #1}\,(#3)\,#2}
\def\pan#1#2#3{{Phys. Atom. Nucl.\,}{\bf #1}\,(#3)\,#2}
\def\app#1#2#3{{Acta. Phys. Pol.\,}{\bf #1}\,(#3)\,#2}
\def\jmp#1#2#3{{J. Math. Phys.\,}{\bf {#1}}\,(#3)\,#2}
\def\cp#1#2#3{{Coll. Phen.\,}{\bf {#1}}\,(#3)\,#2}
\def\epjc#1#2#3{{Eur. Phys. J.\,}{\bf C{#1}}\,(#3)\,#2}
\def\mpla#1#2#3{{Mod. Phys. Lett.\,}{\bf A{#1}}\,(#3)\,#2}

\end{document}